\documentclass[
reprint,
showpacs,
 amsmath,
 amssymb,
 aps,
 pra
]{revtex4-1}
\usepackage[cmyk]{xcolor}
\usepackage{graphicx} 
\usepackage{dcolumn} 
\usepackage{bm} 
\usepackage[utf8x]{inputenc}
\usepackage[T1]{fontenc}
\usepackage{times}
\usepackage{flafter} 
\usepackage{textgreek} 
\usepackage{mathrsfs}
\usepackage{breqn}
\usepackage{capt-of}
\usepackage{wrapfig}
\usepackage{hyperref}
\usepackage{dsfont}
\usepackage{rotating}

\begin{document}
\title{Numerical time-of-flight analysis of the strong-field photoeffect}
\author{V.~A.~Tulsky, and D.~Bauer\\
{\em Institute of Physics, University of Rostock, 18059 Rostock, Germany}}
\date{\today}
\begin{abstract}
Short-time filtering of the photoionization amplitude extracted straight from the numerical solution of the time-dependent Schr\"{o}dinger equation (TDSE) is used to identify dominant pathways that form photoelectron spectra in strong fields. Thereby,  the ``black-box nature'' of TDSE solvers only giving the final spectrum is overcome, and simpler approaches, e.g., semi-classical based on the strong-field approximation, can be tested and improved. The approach also allows to suppress intercycle quantum interference between pathways removing patterns that are usually washed out in experiments.
\end{abstract}
\maketitle

\section{Introduction}

In order to predict photoelectron spectra that are measured in intense laser-matter interaction experiments, theorists proposed a plethora of approaches, from first-principle numerical \cite{qprop3, qprop_tsurff, qprop, Brown_CPC_2020, Patchkovskii_CPC_2016, Majety_NewJPhys_2015, Tao_Scrinzi_NewJPhys_2012} to simplified, semi-classical \cite{Symphony_on_SFA, Miloshevich_PRA_2017, Cohen_PRA_2001, Liu_Springer_2013}.  In general, the simpler and the more analytical a model is, the more insight into the photoionization process it gives while precise quantitative predictions are hardly possible. On the other hand, the most  accurate results are obtained from ab initio solutions of the time-dependent Schr\"{o}dinger equation (TDSE). However, as in an experiment, it is hard to disentangle in TDSE simulations all the various processes the photoelectron undergoes on its way to the detector. The strong-field approximation (SFA) \cite{Keldysh_1965, Faisal_1973, Reiss_1980} took the lead as far as insight into the strongly nonlinear electron dynamics is concerned, especially when formulated in terms of semi-classical quantum orbits \cite{Kopold_OptComm_2000, Salieres_Science_2001, Miloshevich_JPhysB_2006, Popruzhenko_JPhysB_2014, Symphony_on_SFA}. The investigation of these orbits and their weights allows to identify the dominant ionization pathways  \cite{Kopold_OptComm_2000, Salieres_Science_2001, Miloshevich_JModOpt_2006, Yan_PRL_2010, Miloshevich_PRA_2016}. However, photoelectron spectra calculated by such approaches can be quantitatively orders of magnitude off or have a qualitatively wrong shape because of Coulomb effects  \cite{Popruzhenko_JModOpt_2008, Keil_PRL_2016, Maxwell_PRA_2017, Faria_RepProgPhys_2020}. The influence of the Coulomb potential on the photoelectron can be easily accounted for within classic-trajectory Monte-Carlo simulations \cite{Liu_Springer_2013, Hao_PRA_2020}, which, however,  lack interference effects intrinsic to quantum systems.
In the present work, we propose ways to identify the relevant ionization pathways in ab initio solutions to the TDSE. Our method combines the accuracy of the TDSE with the insight offered by quantum-orbit methods. Moreover, simple, semi-analytical models can be benchmarked and improved using the TDSE-based results.

\begin{figure}[h!]
\includegraphics[scale=0.12]{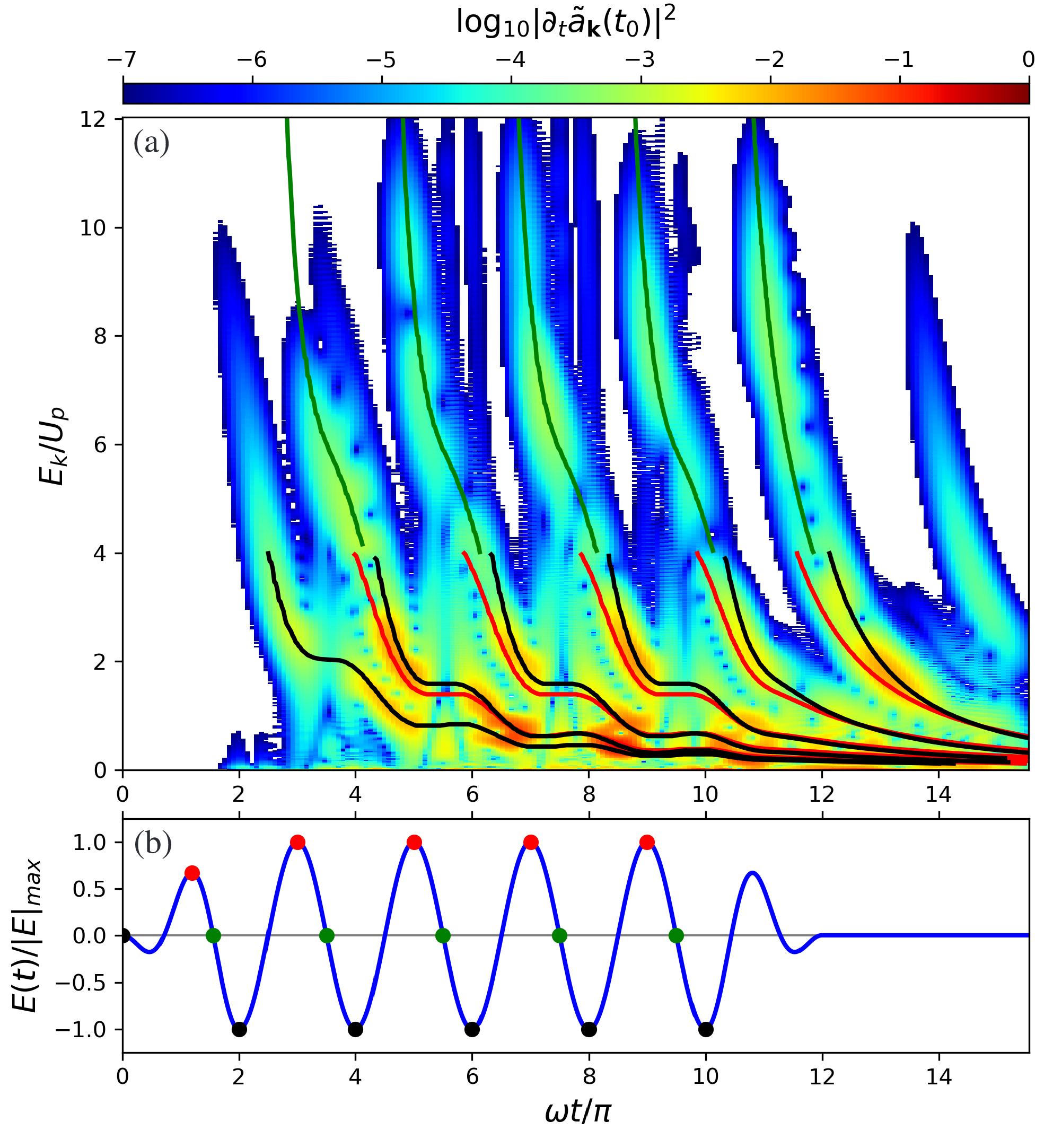}
\caption{(a) Time-energy-resolved photoelectron spectrum from argon atom in the laser propagation direction computed from (\ref{tilde_a_k_T}) and normalized to its maximum. Laser wavelength $\lambda=2000$nm, intensity $I=10^{13}$W/cm$^2$. Flux-capturing surface is at $R=350$ atomic units. For electrons that originate from times $t_{\mathrm{i}}$ when $\vert E(t_{\mathrm{i}}) \vert =E_{\max}$ and that have energies $E_k<4U_{\mathrm{p}}$ the arrival times $t_{\mathrm{reg}}$ according to (\ref{t_reg}) are indicated by red and black lines; for those that originate from times $t_{\mathrm{i}}$ when $E(t_{\mathrm{i}})=0$ and that have energies $E_k>4U_{\mathrm{p}}$ the arrival times are indicated by green lines. (b) Laser profile $E(t)$ with points referring to colored lines in (a).}
\label{fig:pes_tilde_2000}
\end{figure}

We start with a brief introduction to the time-dependent surface flux method (tSURFF) for the calculation of photoelectron spectra \cite{Ermolaev_PRA_1999, Tao_Scrinzi_NewJPhys_2012}. Consider an electron described by the state $\vert \Psi(t)\rangle$ and a laser field ${\bf E}(t)$ defined via its vector potential ${\bf A}(t) = -\int_0^t {\bf E}(t')dt'$ in dipole approximation.  Initially in a bound state of the atomic potential $\vert \psi_{\mathrm{bound}}(t) \rangle$, transitions  to continuum states $\vert {\bf k}(t)  \rangle$ due to interaction with the laser may occur, contributing to the free part $\vert \psi_{\mathrm{free}}(t) \rangle$,
\begin{align}
    \vert \Psi(t)\rangle &= \vert \psi_{\mathrm{bound}}(t) \rangle + \vert \psi_{\mathrm{free}}(t) \rangle, \\
    \vert \psi_{\mathrm{free}}(t)  \rangle &= \int \vert {\bf k}(t)  \rangle \langle {\bf k}(t)   \vert \psi_{\mathrm{free}}(t)  \rangle d{\bf k}.
\end{align}
The momentum-resolved photoelectron spectrum is defined as 
\begin{equation}
    Y({\bf k}) = \vert a_{{\bf k}}(T)\vert^2 = \vert \langle {\bf k}(T)  \vert \psi_{\mathrm{free}}(T)  \rangle \vert^2
\end{equation}
where time $T \to \infty$. Since all bound states are negligible at large enough distances where the photoionized part of the wavefunction is  localized,
\begin{equation}\label{Y_k}
    Y({\bf k}) = \vert a_{{\bf k}}(T)\vert^2 \simeq \vert \langle {\bf k}(T)  \vert  \Theta(r-R) \vert \Psi(T)  \rangle \vert^2
\end{equation}
with the Heaviside step function $\Theta(r-R)$ cutting away the contributions from distances $r<R$. 
With $a_{\bf k}(0)=0$, the amplitudes for the continuum states of momentum ${\bf k}$ at the final time can be written as 
\begin{equation}\label{a_k_T_1}
    a_{\bf k}(T)=\int_0^{T}\partial_t a_{\bf k}(t) dt.
\end{equation}
Using $\partial_t \bullet = i [ \hat{H}(t),\bullet ]$, it is straightforward to show that the amplitudes $a_{\bf k}(T)$ can be expressed as a time integral over the flux ${\bf j}_{{\bf k}}$ through the surface $S_R$ at $r=R$,
\begin{equation}\label{a_k_T_2}
    a_{\bf k}(T) = \int_0^{T}  \left( \int_{S_R} {\bf j}_{{\bf k}}({\bf r},t)d{\bf S}_{{\bf r}} \right) dt
\end{equation}
with the normal vector on the surface $d{\bf S}_{\bf r} = {\bf n}_{\bf r} dS$.
The particular form of the surface flux  ${\bf j}_{{\bf k}}({\bf r},t)$ used in this work is described in the Appendix and in \cite{qprop_tsurff}.

\begin{figure}[h!]
\includegraphics[scale=0.18]{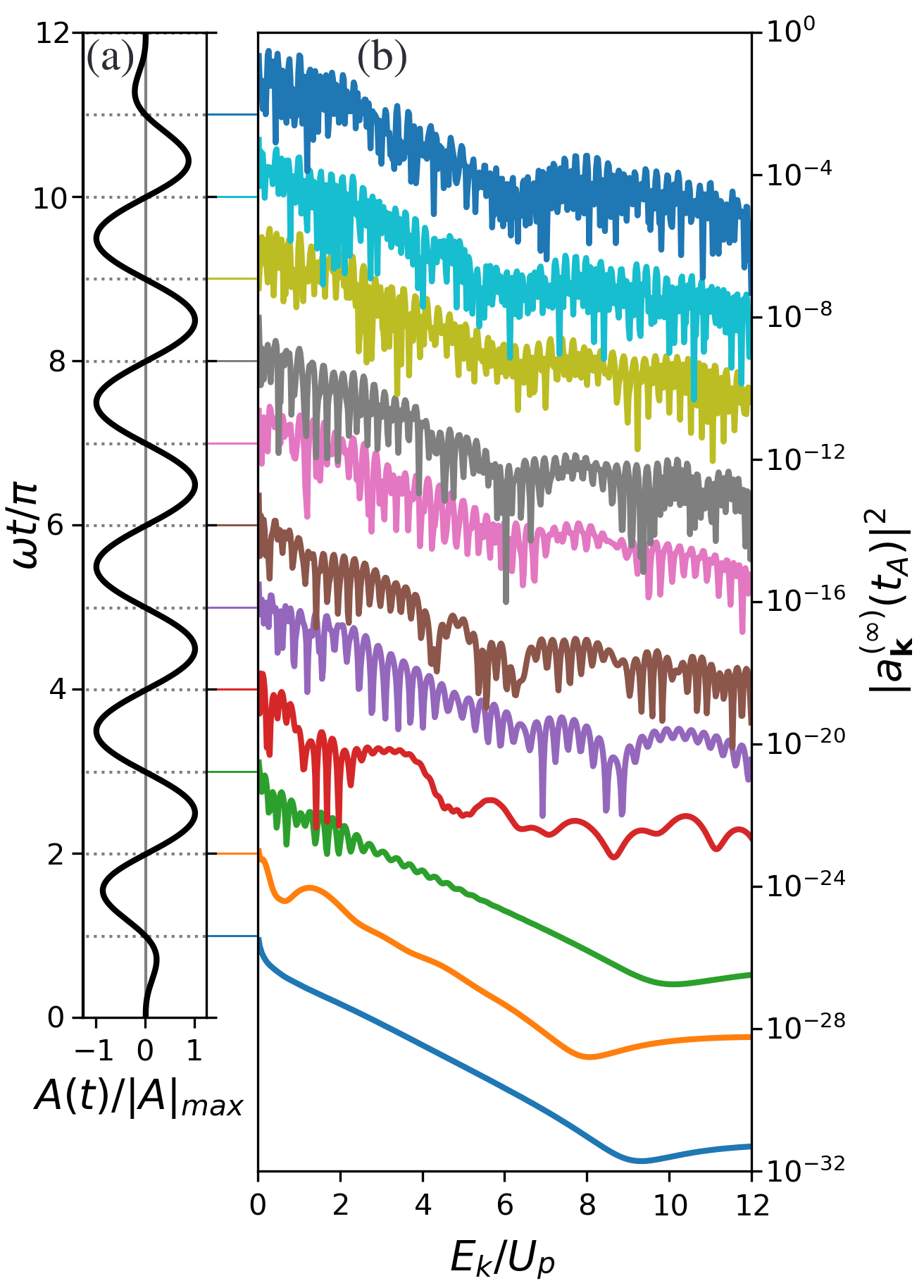}
\caption{(a) Vector potential $A(t)$. (b) Time-energy-resolved photoelectron spectra from argon atom in laser polarization direction with the laser vector potential suddenly switched off at times when it reaches zero. Spectra were calculated using the amplitude (\ref{a_k_inf}). Results are shifted along the vertical axis for better visualization. Laser wavelength $\lambda=800$nm, intensity $I=10^{14}$W/cm$^2$.}
\label{fig:pes_integr_800}
\end{figure}

\section{Time-energy photoelectron distribution}

Before tSURFF became widely used the window-operator method (WOM) \cite{Schafer_PRA_1990} was a common way to obtain the photoelectron spectrum from the final wavefunction calculated by some TDSE solver. While WOM gives the correct total electron spectrum, momentum or angle-resolved spectra are only approximate \cite{Bauke_book_2017, Fetic_PRA_2020}. Nevertheless, phase-space distributions after the laser pulse were obtained using WOM, which proved useful for comparisons with semi-classical theories \cite{Bauer_PRL_2005, Bauer_JModOpt_2006}. The main technical advantage of tSURFF over WOM  is that there is no need to keep track of the full wavefunction, which rapidly spreads over a huge area due to strong ionization. Instead, the photoionization amplitude $a_{\bf k}(T)$ is expressed as a time integral, as mentioned above.  Albeit a mathematical trick in the first place, this provides the opportunity to study the build-up of the photoelectron spectrum in a time-resolved way. In \cite{Serov_PRA_2013}, the integrand itself has been investigated, which, however, is highly oscillatory. Instead, we introduce a Gaussian time window in the integrand, \begin{equation}\label{tilde_a_k_T}
    \partial_t \tilde{a}_{\bf k}(t_0)=\int_{0}^{T} e^{-\frac{(t-t_0)^2}{2t_{\mathrm{w}}^2}} \partial_t a_{\bf k}(t) \frac{dt}{\sqrt{2\pi t_{\mathrm{w}}^2}}, 
\end{equation}
which amounts to the zero-frequency component of a Gabor transform \footnote{In a similar way, the build-up of high-harmonic generation spectra are analyzed (see, e.g., \cite{Chiril_PRA_2010, Murakami_PRA_2013, Surez_PRA_2017, Tancogne_SciAdvances_2018, Chen_EPL_2019})}.
Note that the transform  \eqref{tilde_a_k_T} conserves the total amplitude and the yield because integration over $t_0$ gives back (\ref{a_k_T_1}), and
\footnote{In order to suppress the effects of the finiteness of $T$ a smooth time window can be applied at times approaching $T$ (see Eq.~(\ref{window_T}) in the Appendix). Then one also insures that the flux is zero close and beyond the limits $t_0=0$ and $t_0=T$ and may safely replace the limits $\pm \infty$ for $t_0$ by $0$ and $T$.}
\begin{equation}\label{Y_k_tilde}
    \tilde{Y}({\bf k}) =\left\vert \int_{-\infty}^{\infty} \!\! \left(\int_{0}^{T} \!\! e^{-\frac{(t-t_0)^2}{2t_{\mathrm{w}}^2}} \partial_t a_{\bf k}(t) \frac{dt}{\sqrt{2\pi t_{\mathrm{w}}^2}} \right) dt_0 \right\vert^2 = Y({\bf k}).
\end{equation}
We choose $\omega t_{\mathrm{w}}=0.2\pi$, i.e., a tenth of a laser cycle, so that only intracycle interference is captured at this subcycle time resolution. An example for such a time-resolved spectrum is shown in Fig.~\ref{fig:pes_tilde_2000} for an argon atom in a 6-cycle laser pulse with an ``1-4-1 envelope'' (i.e., ${\bf A}(t)$ has a 4-cycle flat-top central part and an 1-cycle $\sin^2$-ramping on each side). 

When applying the tSURFF method to calculate photoelectron spectra in  TDSE calculations, one usually adopts a classical picture for laser-driven photoelectrons to estimate the time limit $T$ for integration as the time that the slowest electrons of interest need to reach the flux-capturing surface, \begin{equation}\label{Tmin}
    T \geq T_{\mathrm{pulse}}+\frac{R}{k_{\mathrm{min}}}. 
\end{equation}
The density map in Fig.~\ref{fig:pes_tilde_2000} does not only support the classical  estimate \eqref{Tmin} but additionally allows to resolve times when  electrons with a certain energy most likely arrive at the detector. Typically, strong-field photoelectron spectra consist of so-called ``direct'' electrons  with energies $E_k < (2 - 4) U_{\mathrm{p}}$ (where $U_{\mathrm{p}}=I/4\omega^2$ is the ponderomotive energy in atomic units), and ``rescattered'' electrons at energies $E_k<10 U_{\mathrm{p}}$. The direct electrons  are emitted at times with high absolute value of the electric field $\vert {\bf E}(t) \vert$ while the final energy of the rescattered electrons is determined at scattering times  when the absolute value of the vector potential $\vert {\bf A}(t) \vert$ approaches its maximum (we consider first-order returns only) \cite{Kopold_OptComm_2000}. Indeed, from Fig.~\ref{fig:pes_tilde_2000} it is clear that the curves $t_0 = t_{\mathrm{reg}}(k,t_{\mathrm{i}})$ obtained from
\begin{equation}\label{t_reg}
    R = \int_{t_{\mathrm{i}}}^{t_{\mathrm{reg}}} A(t)dt + k(t_{\mathrm{reg}}-t_{\mathrm{i}})
\end{equation}
match the corresponding times of highest time-resolved yield. Note that for  registration times $t_{\mathrm{reg}}>T_{\mathrm{pulse}}$ this equation simplifies so that the   curves $t_0 = t_{\mathrm{reg}}(k,t_{\mathrm{i}})$ are less ``wiggly''. In any case, the time-resolved yield is nicely aligned along these curves of arrival.

\begin{figure}[h!]
\includegraphics[scale=0.14]{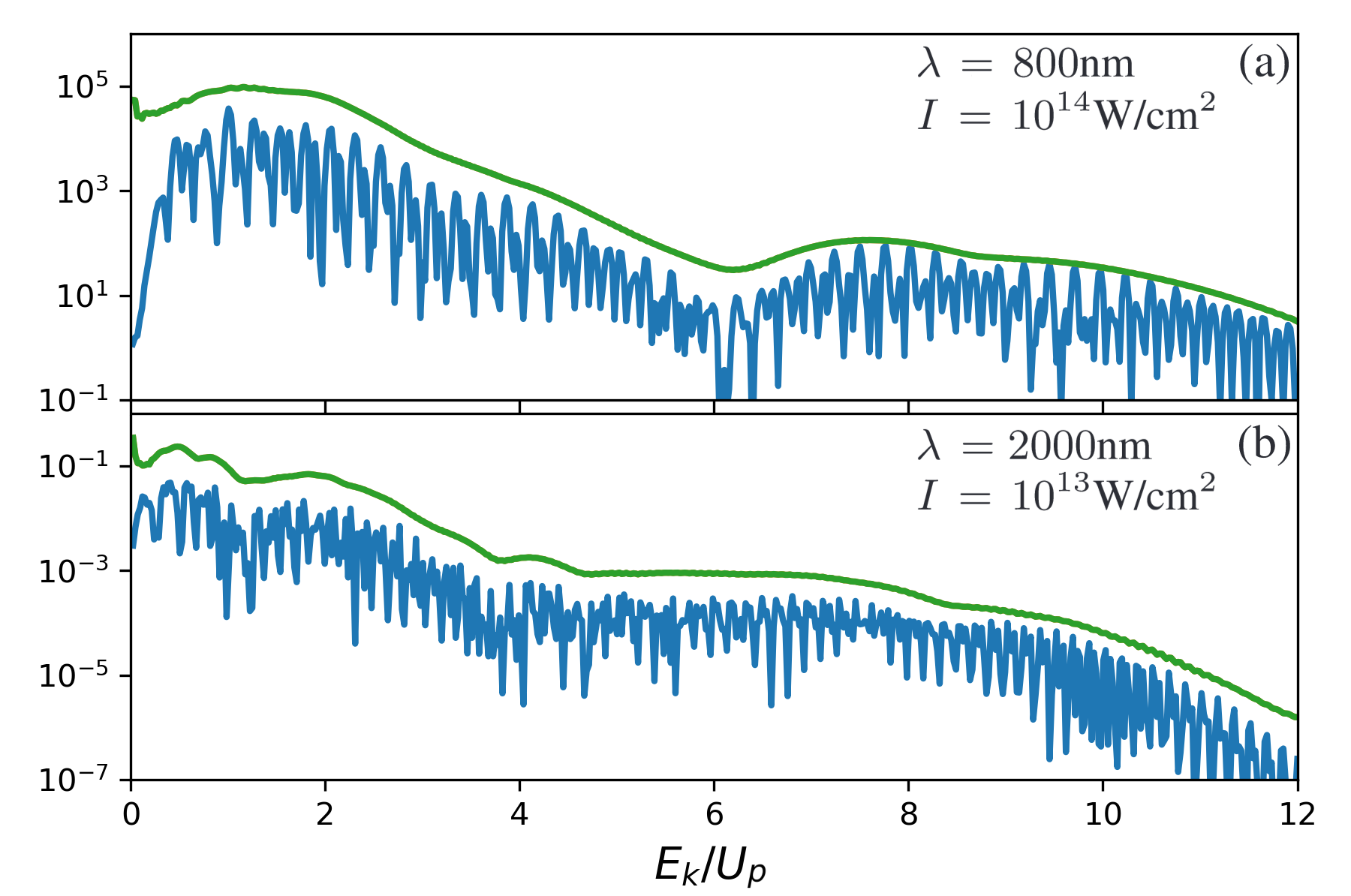}
\caption{Photoelectron spectra in laser polarization direction computed according (\ref{Y_k_tilde}) (blue), according (\ref{Y_k_abs_tilde}) (green). Spectra according  (\ref{Y_k}) coincide with the blue curves up to numerical integration error. Laser pulses with a 1-4-1 envelope (see text) were used in both examples with (a) $\lambda=800$nm at $I=10^{14}$W/cm$^2$ ($U_{\mathrm{p}}=3.85\omega$) and  (b) $\lambda=2000$nm at $I=10^{13}$W/cm$^2$ ($U_{\mathrm{p}}=6.02\omega$).}
\label{fig:PES_trend}
\end{figure}

A way to visualize the build-up of intercycle interference and the rescattering plateau was proposed in \cite{Borbely_PRA_2019}. At times $t_E$ where $ {\bf E}(t_E) =0$, the contribution $\vert \psi_{\mathrm{bound}}(t_E)\rangle$ was projected out  of the state  $\vert \Psi(t_E)\rangle$. The calculated spectra were then equal to those where the laser pulse was suddenly terminated at times $t_E$ because the authors worked in the length gauge and, thus, eigenstates of their Hamiltonian at times $t_E$ were eigenstates of the laser-free Hamiltonian. Analogously, within the velocity gauge this approach is valid at times $t_A$ such that ${\bf A}(t_A)=0$. After obtaining $\vert \Psi(t_A)\rangle$ one may set ${\bf A}(t>t_A)=0$ and
 apply the iSURFV method to efficiently post-propagate according to the field-free Hamiltonian $\hat{H}_0$  \cite{Serov_PRA_2013, Morales_JPhysB_2016, qprop3},
\begin{equation}\label{a_k_inf}
    a^{(\infty)}_{\bf k}(t_A) = a_{\bf k}(t_A)+\delta a_{\bf k}^{(\infty)}(t_A)
\end{equation}
with
\begin{equation}\label{delta_a_k_inf}
    \delta a_{\bf k}^{(\infty)}(t_A) = \int_{S_R} {\bf J}^{(\infty)}_{{\bf k}}({\bf r},t_A)d{\bf S}_{{\bf r}}.
\end{equation}
The particular form of ${\bf J}^{(\infty)}_{{\bf k}}({\bf r},t_A)$ used is given in the Appendix and \cite{qprop3}. Figure~\ref{fig:pes_integr_800} shows how intercycle interference and the rescattering plateau form during the laser-atom interaction. Such intermediate spectra, of course, suffer from gauge non-invariance but nevertheless may be useful to study because similar build-ups are observed in SFA-based theories where the photoelectron amplitude $M_{\bf k}$ is also given by a time integral. In quantum-orbit theory, this time integral is evaluated using the saddle-point method, and the saddle points can be interpreted as the complex ionization times $t_s$ that contribute predominantly,
\begin{equation}\label{a_SFA}
    a_{\bf k}^{\mathrm{SFA}}(T) = \sum_s M_{\bf k}(t_s).
\end{equation}
A sudden shutdown of the laser in the TDSE simulations corresponds to discarding ionization times with Re$(t_s)>t_E$ (or Re$(t_s)>t_A$ in velocity gauge) \cite{Kopold_OptComm_2000, Li_OptExpr_2016, Han_PRL_2017, Nayak_PhysRep_2019}.

Continuing exploring features accessible in SFA-based theories, we address the possibility to exclude intercycle interference (see, e.g., \cite{Xie_PRL_2017, Nayak_PhysRep_2019}).
In quantum orbit theory, this can be achieved by considering an incoherent sum over saddle points,
\begin{equation}\label{Yield_SFA_incoh}
    Y_{\mathrm{abs}}^{\mathrm{SFA}}({\bf k}) = \left( \sum_s \vert M_{\bf k}(t_s)\vert \right)^2.
\end{equation}
This trick is used to visualize trends in photoelectron distributions generated by long laser pulses where otherwise intercycle interference could suppress the signal apart from positions of the so-called above-threshold-ionization (ATI) peaks
\begin{equation}
    E_k^{\mathrm{ATI}} = n \omega - I_{\mathrm{p}} - U_{\mathrm{p}}, \qquad n \geq n_{\mathrm{min}}.
\end{equation}
The presence of ATI peaks in TDSE spectra may complicate the comparison with experimental results where ATI peaks are washed out because of focal averaging, for instance. We propose a way to eliminate ATI peaks from TDSE spectra using the modulus of the time-averaged integrand
\footnote{Note, that the SFA time integral (sum over saddle points in Eq.~(\ref{a_SFA})) is taken over ionization, not registration times. Therefore, Eq.~(\ref{Yield_SFA_incoh}) does not directly match with Eq.~(\ref{Y_k_abs_tilde}) where integration over registration times is performed. Nevertheless, both tricks successfully remove the pattern of the ATI peaks from spectra.}
,
\begin{equation}\label{Y_k_abs_tilde}
    \tilde{Y}_{\mathrm{abs}}({\bf k})=\left( \int_{0}^{T}\left\vert\int_{0}^{T} e^{-\frac{(t-t_0)^2}{2t_{\mathrm{w}}^2}} \partial_t a_{\bf k}(t) \frac{dt}{\sqrt{2\pi t_{\mathrm{w}}^2}} \right\vert dt_0 \right)^2.
\end{equation}
Figure~\ref{fig:PES_trend} shows spectra calculated in this way. The total number of ATI peaks in the $E_k<10U_{\mathrm{p}}$ domain scales  $\sim I \lambda^3$. As a consequence, a higher energy resolution is required for longer wavelengths if ATI peaks need to be resolved. Our method to eliminate the ATI peaks allows to study more efficiently trends in spectra as a function of, e.g.,  laser wavelength or intensity.

\section{Time and angle-resolved electron distributions}

\begin{figure}
\includegraphics[scale=0.41]{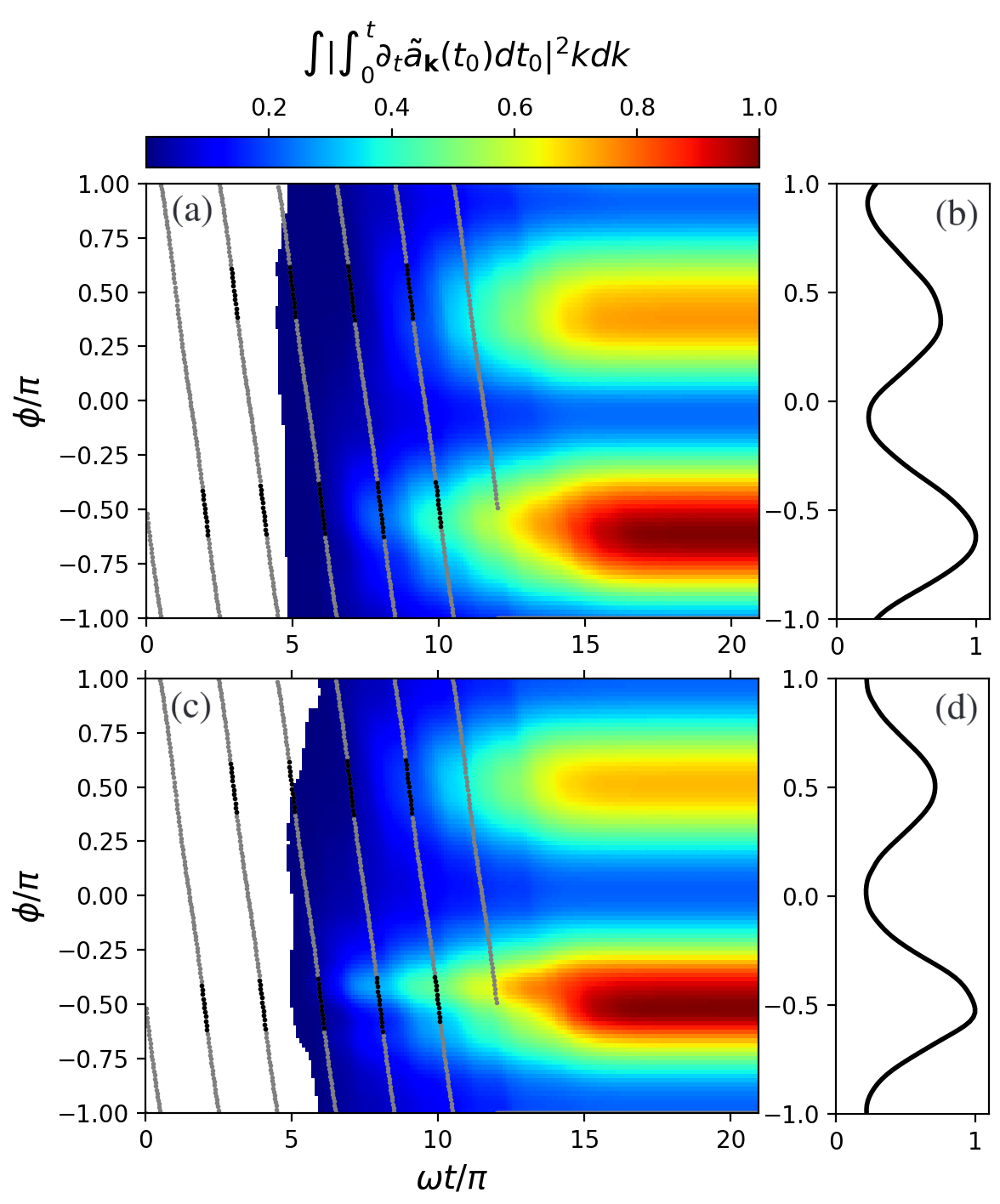}
\caption{(a,c) Integrated flux according  (\ref{Y_t_phi}) vs angle and time for a hydrogen in a laser pulse with wavelength $\lambda=800$nm, intensity $I=10^{14}$W/cm$^2$, ellipticity $\epsilon=0.882$. Flux-capturing surface is put at $R=200$. Gray lines correspond to angles of $-{\bf A}(t)$, black lines on top of them indicate angles where $\vert {\bf E}(t)\vert> 0.986 E_{\mathrm{max}}$. (b,d) Total angular distributions, i.e., sections through $t=T$ of (a,c). (a,b) for hydrogenic binding potential, (c,d) for short-range potential with same ionization potential $I_{\mathrm{p}}=13.6$eV.}
\label{fig:elliptic}
\end{figure}

\begin{figure}
\includegraphics[scale=0.41]{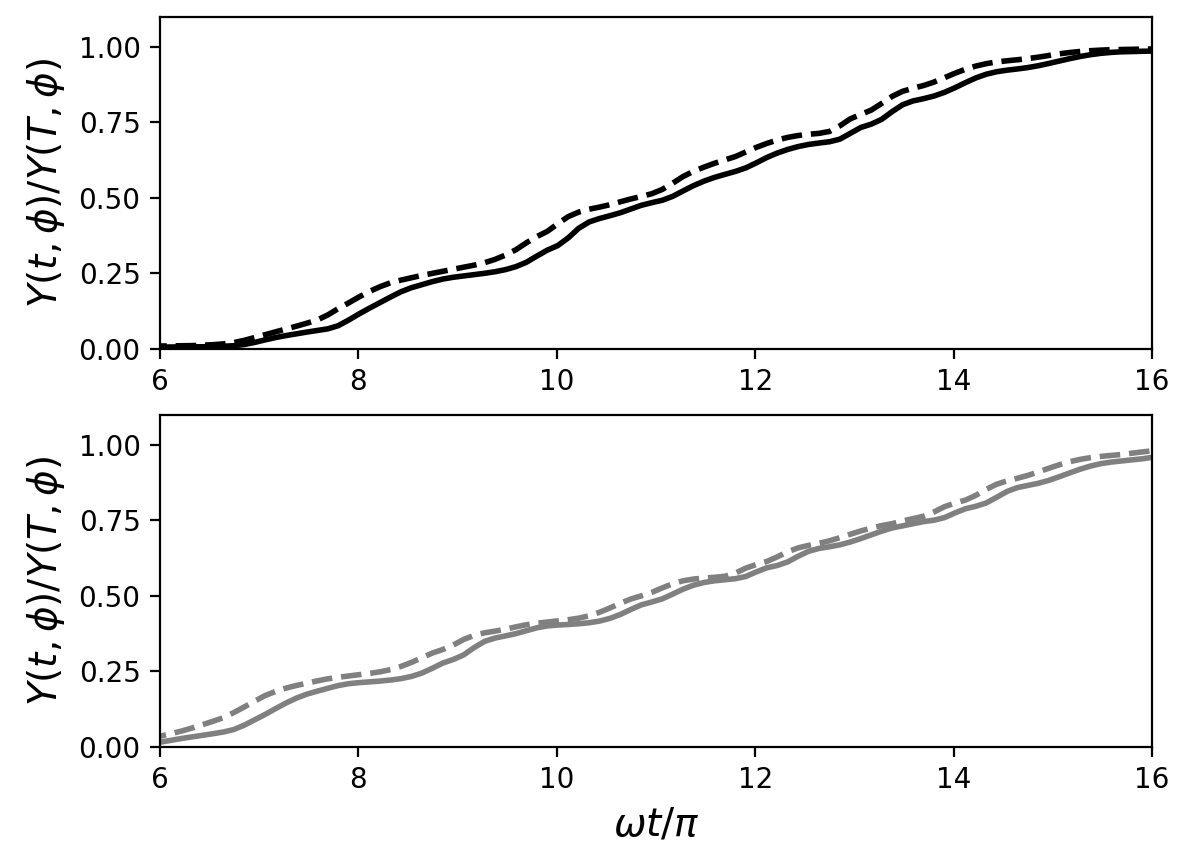}
\caption{A section through the time and angle resolved distributions shown in  Fig.~\ref{fig:elliptic} at angles $\phi$ where the total yield is maximum, i.e.,  $\phi=-0.616\pi$ (solid black) and $\phi=0.373\pi$ (solid gray) for hydrogen, $\phi=-0.495\pi$ (dashed) and $\phi=0.495\pi$ (dashed gray) for short-range potential. All four curves are normalized to their maxima.}
\label{fig:elliptic_cut}
\end{figure}

The methods introduced so far are also applicable to elliptically polarized laser pulses
\begin{equation}
    {\bf A}(t) = A_0(t) {{~ ~\sin(\omega t)}\choose{\epsilon\cos(\omega t)}}.
\end{equation}
We use the same 1-4-1 envelope $A_0(t)$ as in the previous examples and laser parameters similar to \cite{Guo_ArXiv_2019}: wavelength $\lambda = 800$nm, intensity $I=10^{14}$ W/cm$^2$, and ellipticity $\epsilon=0.882$. In Fig.~\ref{fig:elliptic}, we show the angular photoelectron distribution as a function of time $t$ until the accumulated yield does not change anymore. Only electrons in the polarization plane $\theta=\pi/2$ are considered,
\begin{equation}\label{Y_t_phi}
    Y(t,\phi) = \int \left\vert \int_0^t  \partial_t \tilde{a}_{\bf k}(t_0) dt_0\right\vert^2 kdk.
\end{equation}
According to  the semi-classical picture, the yield should be highest at angles $\phi \sim \pm \pi/2$ where 
\begin{equation}
    -{\bf A}(t_A) = \vert {\bf A}(t_A)\vert {{\cos\phi}\choose{\sin\phi}}
\end{equation}
holds, and  $t_A$ are the times of maximum  $\vert {\bf E}(t)\vert$. The mean momentum in atomic units at those angles is
\begin{equation}
    \langle k(\phi) \rangle = \frac{\int  k Y(k,\phi) kdk}{\int  Y(k,\phi) kdk} \simeq 0.8. 
\end{equation}
We put our flux-capturing surface at $R=200$  so that the surface flux should accumulate at times
\begin{equation}
    t_{\mathrm{reg}} = t_A(\phi) + \frac{R}{ \langle k(\phi) \rangle} \sim t_A + 2.5 \cdot 2\pi/\omega.
\end{equation}
Figures~\ref{fig:elliptic}(c,d) show that this simple estimate is confirmed for a short-range binding potential (defined in the Appendix) while it is violated for a hydrogenic binding potential (see Figs.~\ref{fig:elliptic}(a,b)) where a deviation in the arrival time $\omega\Delta t_{\mathrm{reg}} =-\Delta \phi \sim 0.12\pi$  is observed and illustrated in Fig.~\ref{fig:elliptic_cut}. The mapping of registration, ionization or tunneling-time delays to  angular shifts is the basic idea behind so-called  ``attoclock''   experiments \cite{Eckle_Nature_2008, Pfeiffer_ChemPhys_2013, Landsman_Optica_2014, Han_PRL_2019}.

\section{Conclusions}

We showed how the dominant pathways that constitute  photoelectron spectra in strong-field laser ionization experiments are revealed by applying a short-time filter to the ionization amplitude that is calculated anyway within the efficient tSURFF approach incorporated in state-of-the-art strong-field TDSE solvers. Our method allows to connect ab initio TDSE simulations with simple and intuitive semi-analytical theories, thus providing insight and a way to benchmark and improve simple models. We also pointed out an efficient way to remove intercycle interference, revealing the envelopes of photoelectron spectra without ATI peaks.
 
 Future work may concentrate on regimes where simple, semi-classical theories actually do {\em not} work, e.g., in the over-barrier-ionization regime or if excited states play a role.
 
\section*{Acknowledgement}

This work was supported by the project BA 2190/10 of the German Science Foundation (DFG).

\section*{Appendix} 
All TDSE solutions in the present paper were obtained with the QPROP software \cite{qprop,qprop_tsurff,qprop3}. For the examples with linear polarization an argon pseudo potential  
\begin{equation}
    U_{\mathrm{Ar}}(r)=-\frac{1+17e^{-c r}}{r}, \qquad r<R_{\mathrm{co}}
\end{equation}
with $c=2.2074$ was used.  This leads to an ionization potential of  $I_{\mathrm{p}}=15.8$eV for the $3p$ initial state on a radial grid of resolution $dr=0.1$. The flux-capturing surface was put at $R=350$. 

For the example  with elliptical polarization the hydrogenic potential
\begin{equation}
    U_{\mathrm{H}}(r)=-\frac{1}{r}, \qquad r<R_{\mathrm{co}}
\end{equation}
and a short-range potential
\begin{equation}
    U_{\mathrm{sr}}(r)=-C_1\frac{e^{-c_2 r}}{r}, r<R_{\mathrm{co}}
\end{equation}
with $C_1=5.2074$ and $c_2=5.0$ were used. Both lead to a ground-state ionization potential  $I_{\mathrm{p}}=13.6$eV. The flux-capturing surface was put at  $R=200$. In all calculations the long-range Coulomb tail was removed by matching it at  $r=R_{\mathrm{co}}=50$ to a linear roll-off that reaches zero at $r=2R_{\mathrm{co}}$. Since $R>2R_{\mathrm{co}}$, this approximation allows to use Volkov functions for $\vert {\bf k}(t)\rangle$ and the Volkov Hamiltonian 
\begin{equation}
    \hat{H}(t)=\frac{(-i\nabla+{\bf A}(t))^2}{2}-\frac{ A^2(t)}{2},
\end{equation}
(with the purely time-dependent  $A^2(t)/2$ term transformed away), leading  
 to the density flux 
\begin{multline}
    {\bf j}_{{\bf k}}({\bf r},t) = \frac{i}{2} \langle {\bf k}(t)  \vert \nabla+i{\bf A(t)}\vert {\bf r}\rangle \langle {\bf r} \vert \Psi(t)\rangle \\- \frac{i}{2} \langle {\bf k}(t)  \vert {\bf r}\rangle\langle {\bf r} \vert \nabla+i{\bf A(t)}\vert \Psi(t)\rangle
\end{multline}
needed in equation (\ref{a_k_T_2}).
Unless the iSURFV method was used, the upper limit for the time integration was chosen $T=T_{\mathrm{pulse}}+500$. The time window
\begin{equation}\label{window_T}
    w(t) = 1 -e^{-\frac{(t-T)^2}{2T_{\mathrm{w}}^2}}
\end{equation}
with $T_{\mathrm{w}}=100$ was added to suppress the effects of finiteness of $T$.
In the expression (\ref{delta_a_k_inf}), where we use the iSURFV approach \cite{Serov_PRA_2013, Morales_JPhysB_2016, qprop3}, the time-integrated flux density is
\begin{multline}
    {\bf J}^{(\infty)}_{{\bf k}}({\bf r},t_A) =  \frac{i}{2} \langle {\bf k}(t_A)  \vert \nabla\vert {\bf r}\rangle \langle {\bf r} \vert \frac{1}{E_k-\hat{H}_0} \vert\Psi(t_A)\rangle \\- \frac{i}{2} \langle {\bf k}(t_A)  \vert {\bf r}\rangle\langle {\bf r} \vert \nabla\frac{1}{E_k-\hat{H}_0}\vert \Psi(t_A)\rangle. \label{eq:Iinfinity}
\end{multline}

\bibliography{biblio_big}

\end{document}